\documentclass[a4paper]{PoS}

\usepackage{graphicx,amssymb,amsbsy,amsfonts,amssymb,amsmath}
\usepackage{multirow}
\usepackage{stackrel}
\usepackage{cancel}

\newcommand{\be}{\begin{equation}}
\newcommand{\ee}{\end{equation}}

\newcommand{\bea}{\begin{eqnarray}}
\newcommand{\eea}{\end{eqnarray}}

\newcommand{\hm}{\widehat{m}}

\newcommand{\msp}{\hspace{-0.2cm}}
\newcommand{\mmsp}{\hspace{-0.4cm}}

\def\sect#1{section~{\ref{#1}}}

\def\fig#1{fig.~{\ref{#1}}}

\def\tN{\tilde N}

\def\trho{\tilde{\rho}}

\def\talpha{\tilde{\alpha}}

\def\td{\tilde{d}}
\def\hd{\hat{d}}
\def\tu{\tilde{u}}

\def\ta{\tilde{a}}

\def\tell{\tilde{\ell}}
\def\hell{\hat{\ell}}
\def\tm{\tilde{m}}
\def\tn{\tilde{n}}

\def\tpartial{\tilde{\partial}}

\def\tp{\tilde{p}}
\def\hp{\hat{p}}

\def\tl{\tilde{l}}

\def\invprops{\rho^0 \cdots \trho^{(\tN-1)}}

\def\dropinvprops{\rho^0 \cdots \bcancel{\,\,\rho^i} \cdots  \trho^{(\tN-1)}}
\def\measure{[d\rho,d\alpha]}
\def\cutmeasure{[d\alpha]}
\def\rhomeasure{[d\rho]}
\def\cut{\overset{{\rm cut}}{\longrightarrow}}
\def\ibp{IBP}

\title{Towards a Numerical Unitarity Approach for Two-loop Amplitudes in QCD}
\ShortTitle{Numerical Unitarity at Two Loops in QCD}
\author{\speaker{Harald Ita}\thanks{I would like to thank the organisers for the opportunity to present this work.}\\
Physikalisches Institut, Albert-Ludwigs-Universi\"at Freiburg
D-79104 Freiburg, Germany
\\
E-mail: \email{harald.ita@physik.uni-freiburg.de}}
\abstract{
    The numerical unitarity approach has been important for obtaining reliable
    QCD predictions for the LHC.  Here I discuss the extension of the approach
    beyond the leading quantum corrections for computing multi-loop amplitudes.
    The numerical unitarity approach requires a suitable
    parametrization of the loop integrands as a sum of terms that integrate to
    zero (surface terms) and master integrands. 
    The construction and classification of suitable surface terms which match
the propagators structures of Feynman amplitudes is the main technical advance
discussed. A number of spin-offs for integral reduction and further formal
questions are briefly reviewed.  }

\FullConference{Loops and Legs in Quantum Field Theory\\
		24-29 April 2016\\
		Leipzig, Germany}

\begin{document}

\section{Introduction}
\label{IntroductionSection}

During the next decade the experiments at the Large Hadron Collider (LHC) will
deliver measurements of the S-matrix at the 5\% level. Theory predictions with
comparable quality are challenging and time consuming to obtain.  The long-term aim of the
presented research is to provide automated theory tools for fixed-order
computations in QCD in order to match the experimental precision.  Here I
discuss the new methods published in ref.~\cite{IntDec} for the computations of two-loop
scattering amplitudes required for next-to-next-to leading order (NNLO)
predictions.

In general terms, any contribution that increases the breath, depth, precision
or completeness of predictions is important.  Breadth or variety in the final
states facilitates indirect detection and  characterization of new physics.
Depth towards high-multiplicity final states allows to probe the kinematic
dependence of interactions. Finally, the percent-level measurements of Standard
Model (SM) cross sections requires precision predictions often including
second-order quantum corrections.  Although challenging to obtain phenomenological
amplitudes have an increased value during the run-time of the LHC making
multi-loop computations worthwhile in many ways. 

During the last ten years a similar challenge was solved by the theory
community. The automation of NLO computations in the QCD and electroweak
coupling
expansion~\cite{VBFNLO,BlackHat}
lifted the understanding of the SM to the 10\% level.
Presently a limited number of NNLO predictions are available with two
final-state objects \cite{2gamNNLO}.
Going beyond this multiplicity or increasing the number of scales requires
advances of the state-of-the-art techniques.
I follow one of the successful approaches for NLO computations, the unitarity
approach \cite{UnitarityI,GenUnitarityIII} and lay
out an extension to multi-loop amplitudes.  In particular, I focus on the
variant of the unitarity methods well suited for numerical computations
\cite{OPP,NumUnitarity,BlackHat}.

This work follows the current advances with amplitude computations exploiting
an algebraic and geometric perspective.  
In particular, the properties of the surfaces defined by vanishing loop
propagators are exploited to guide integral reduction \cite{LZ} and the
unitarity methods (see e.g. \cite{Cycles}) potentially reducing complexity
bottlenecks.

%
\section{The role of unitarity and surface terms}
\label{Notation}

\begin{figure}[t] \center\includegraphics[clip,scale=0.30]{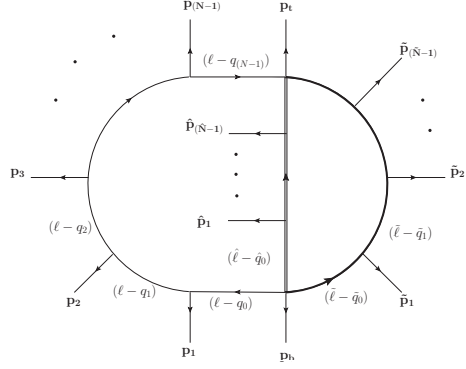}
\caption{
    A generic two-loop integral topology is displayed with the naming
conventions used in the main text. In order to reuse structures well known at
one-loop level one interprets the two-loop topology as three strands. The strands
carry loop momentum $\ell$, $\hell$ and $\tell$, and have external momenta
$p_i$, $\hp_i$, $\tp_i$, $p_t$ and $p_b$ exiting. }
\label{2loopFigure} \end{figure}

Amplitude computations relate Feynman amplitudes to
a universal ansatz in terms of master integrals and their coefficients,
\bea\label{mastereqn}
 {\cal A}^{\rm 2-loop}&:=&
 \int d^D\ell d^D\tell \,\,A^{\rm 2-loop}_{\rm Feyn.}(\ell,\tell)=
\mmsp\sum_{\mbox{\tiny $i$ $\in$ integral basis}} \mmsp  d_i {\cal I}_i\,,\quad
\mbox{and} \quad {\cal I}_i:=\int d^D\ell d^D\tell\,
\frac{m_i(\ell,\tell)}{\invprops}\,.
\eea
Here ${\cal I}_i$ denote master integrals defined by numerator insertions
$m_i(\ell,\tell)$ and the topologies \fig{2loopFigure}.
I focus the discussion on 2-loop amplitudes although the discussion is
valid beyond this. The independent loop momenta are denoted by $\ell$ and
$\tell$, external momenta by $p_i$ and inverse propagators are denoted by
$\rho$-variables, e.g.  $\rho^i=(\ell- q_i)^2-m_i^2$ with $q_i=\sum_j p_j$.

The role of unitarity methods is two-fold; on the one hand the equations
(\ref{mastereqn}) can be split up in to a fine-grained set using analytic
properties of amplitudes (see e.g.~\cite{DOlive}). On the other hand, the
left-hand side of the master equation is simplified to tree-level on-shell
expressions by cutting~\cite{Cutkosky},
\bea\label{unitaritymaster} {\cal A}^{\rm 2-loop}|_{{\rm cut}_{(t,{\cal C})}}=
 \oint_{\cal C} \mu_t[d\alpha]_t \,
\prod_{\mbox{\tiny $i$ $\in$ corners}} A_i^{\rm tree}(\ell,\tell)
=\msp\sum_{\mbox{\tiny $j$ $\in$ cut topologies}} \msp d_j
{\cal I}_j|_{{\rm cut}_{(t,{\cal C})}}+\,\mbox{parent contributions}\,.
\eea
The variable $t$ labels the cut topologies (specified by the list of cut
propagators), $\mu_t$ refers to the integration measure after cutting and
$\alpha$'s denote the remaining integration variables.
Eqn.~(\ref{unitaritymaster}) holds up to `parent contributions' which refers
to integrals with additional (un-cut) propagators.  When solving
eqn.~(\ref{unitaritymaster}) iteratively in the number of cut propagators
starting from cutting the maximal number of propagators, the parent
contributions are already available when needed. Thus,
eqn.~(\ref{unitaritymaster}) are linear equations for the coefficients $d_i$.
As an improvement, one solves for the coefficients $d_i$ directly using
integration contours ${\cal C}_i$ dual to the master integrals
\cite{MLoopContours},
\bea\label{dualcontour} {\cal I}_i|_{{\rm cut}_{(t,{\cal C}_j)}}=\oint_{{\cal
C}_j} \mu_t[d\alpha]\, m_i(\alpha)=\delta_i^j\,, \eea
as manifest by inserting eqn.~(\ref{dualcontour}) into
eqn.~(\ref{unitaritymaster}). The integration over multi-dimensional contours
${\cal C}$ are challenging which motivates the algebraic variant of the
unitarity approach, described next.

Given a basis of integrands (as opposed to the integrands of master integrals)
the algebraic relation, 
\bea \label{integrandmaster}
A^{\rm 2-loop}_{\rm Feyn.}(\ell,\tell)=\msp\sum_{\mbox{\tiny $i$ $\in$
integrand basis}} \frac{ \td_i \,\tm_i(\ell,\tell)}{\invprops} \,,
\eea
holds with yet to be determined coefficient functions $\td_i$.
Such integrand parametrizations have been developed in the recent years
\cite{MLoopParamMO,MLoopParamBFZ,MLoopAlgGeo}.

The unitarity equations emerge, when solving the equations for the
coefficients $\td_i$.  As above, one solves the
eqn.~(\ref{integrandmaster}) topology by topology using their propagator
structure. One proceeds iteratively in the number of propagators starting from
the maximal number of propagators. One considers each topology modulo
expressions with fewer propagators ($\rho^i\sim0$).  Geometrically this amounts
to computing factorization limits which put loop-particles on-shell
\cite{NumUnitarity}. The Feynman amplitude factorizes into trees yielding,
\begin{eqnarray} \label{numunitarity}
\mmsp\lim_{(\ell,\tell)\rightarrow cut_t \,value}{}\mmsp A^{\rm 2-loop}_{\rm
Feyn.}(\ell,\tell)= \prod_{\mbox{\tiny $i$ $\in$ corners}} A^{\rm
tree}_{i}(\ell,\tell)|_{{\rm cut}_t} = \mmsp\sum_{ \mbox{\tiny $i$ $\in$
numerator basis}} \msp\td^i\, \tm_i(\ell,\tell)|_{{\rm cut}_t} + \mbox{parent
contributions}.
\label{ampldecomposition} \end{eqnarray}
Technically the on-shell limits are identical to cutting but with the
integration omitted.  The linear equations for $\td_i$ are algebraic and
suited for numerical implementation.  However, in order to obtain the loop
amplitude the integrand basis has to be reduced to master integrals.

Here I discuss a more direct approach~\cite{IntDec} inspired by one-loop
computations~\cite{OPP,NumUnitarity}. One decompose the integrand, i.e. the
right-hand side of eqn.~(\ref{integrandmaster}) directly into master integrands
(associated to master integrals) and vanishing integrals (surface terms),
\begin{eqnarray}
A^{\rm 2-loop}_{\rm Feyn.}(\ell,\tell) &=&\sum_{ \mbox{\tiny $i$ $\in$ master
integrals}} \frac{d^i\, m_i(\ell,\tell)}{\invprops} + \sum_{ \mbox{\tiny $j$
$\in$ surface terms}} \frac{\hd^j \hm_j(\ell,\tell)}{\invprops}\,,
\label{tensordecomposition}
\end{eqnarray} 
with the properties,
\begin{eqnarray}\label{surfaceterms} {\cal S}_j=\int d^D\ell d^D\tell \frac{
\hm_j(\ell,\tell)}{\invprops}=0\,.  \end{eqnarray}
The algebraic equations (\ref{numunitarity}) hold just as before and yield
$d_i$ and $\hat d_j$. Thus one obtains directly the coefficients $d_i$ of the
master integrals.  The coefficient of the surface terms
$\hd_j$ are but required as auxiliary information when solving the algebraic
equations (\ref{tensordecomposition}).

An integrand representation in terms of surface terms and master integrands
(\ref{tensordecomposition}) has been suggested only recently~\cite{IntDec}.
Naively, integrands of \ibp{} identities have all the properties
needed, however, for unitarity cuts one requires very
specific propagator powers. This obstruction is overcome in the
construction of \ibp-relations using a particular choice of `\ibp-generating'
vectors~\cite{KosowerIBP}.
I follow the methods developed in ref.~\cite{IntDec} to explicitly
construct and classify such \ibp-generating vectors and obtain the surface terms
topology by topology.

\section{Heuristic picture for loop integrals}
\label{adaptedcoordinates}

In order to clarify important mathematical structures, I start with a general
discussion of loop integrals.
A general coordinate change from the loop-momenta
($\ell^\mu$ and $\tell^\mu$) to inverse propagators $\rho^i$ as integration
variables~\cite{DOlive} make the geometric structures of loop integrals manifest.
Given the miss match in the number of propagators and loop-momentum
components additional angular variables, $\alpha$'s have to be introduced.  If
propagators are dependent on each other the integrals are reducible
algebraically (using Gram determinants) and do not have to be considered.

The integration variables of the loop integrals are the inverse propagators and the $\alpha$-coordinates,
\begin{eqnarray}
    {\cal I}_i=\int \frac{\rhomeasure}{\invprops} \,\times\, \tilde m_i
    \,\mu_t \cutmeasure_t\,,  \label{motivationcohomology}
\end{eqnarray}
where $\tilde m:=\tilde m(\ell,\tell)$ denotes a numerator insertion, $\mu_t$ is the
non-trivial integration measure from the coordinate change. The expression
$\rhomeasure$ and  $\cutmeasure_t$ denote the differentials of the integration
variables; $\rho$'s and $\alpha$'s respectively. The integration makes a
foliation of loop-momentum space manifest; for fixed propagator values one
obtains an internal geometry parametrized by the $\alpha$-variables. The
unitarity-cut phase space is one such internal slice. 

It is instructive to consider first the integration over an internal space with
the inverse propagators held fixed. One can now use the properties of the
internal space to organize the computation. First of all, one can omit terms
proportional to inverse propagators, since these will be captured by integrals
with fewer propagators.
For simplicity I focus on one-loop integrals next for which the $\alpha$-integration
is performed over a sphere $\sum (\alpha^a)^2=C(\rho)$
\cite{IntDec,NumUnitarity,RevUnitarity}~\footnote{Typically these spheres are
part of the complex internal spaces which are tangent bundles of the real
spheres $TS_d$.}. One can think of the function $\tilde m(0,\alpha)$ being
expanded into spherical harmonics. Only the constant function gives a
non-vanishing integral, while the higher harmonics integrate to zero. The
latter can be interpreted as surface terms with 
the \ibp{}-generating vectors being rotation generators in $\alpha$ space.
Thus one is left
with the non-vanishing integral with a constant numerator.  \ibp-generating
vectors are tangent vectors of the internal space.
In this way coordinates adapted to the propagator structures are a valuable
tool for identifying reduction steps corresponding to canceling propagators and
identifying vanishing integrals.

\section{Classifying surface terms}
\label{ibpvectors}

Surface terms suited for perturbative computations are obtained from total derivatives 
\begin{eqnarray}\label{ibpsurfaceterms}
    \int d^D\ell d^D\tell \left[
    \partial_\mu \left(\frac{ u^\mu\,  \tilde m } {\invprops}\right) + 
    \tpartial_\nu \left(\frac{ \tu^\nu \, \tilde m } {\invprops}\right)\right] =0\,,
\end{eqnarray}
with polynomial vector fields $\{u^\mu, \tilde u^\nu\}$ and numerators $\tilde m:=\tilde m(\ell,\tell)$. 
Propagators powers can be controlled by using `\ibp-generating vector-fields'
fulfilling the equations \cite{KosowerIBP},
\begin{eqnarray} 
    (u^\mu \partial_{\mu} + \tu^\nu
    \tpartial_{\nu}) \,\rho^i    = f^i(\ell,\tell) \,\rho^i\,, \label{fterms}
\end{eqnarray}
for all inverse propagators. 

The relation (\ref{fterms}) has an interesting on-shell interpretation.
When specializing to the on-shell phase spaces with $\rho^i=0$,
the vector fields $\{u^\mu,\tu^\nu\}$ turn into tangent vectors along the
maximal-cut phase spaces.  It is often helpful to construct first the tangent
vectors to the unitarity cut phase spaces and then take them off-shell in a
second step. 

In adapted coordinates the construction of \ibp{}-generating vectors simplifies~\cite{IntDec},
\begin{eqnarray}  \label{ibpcondition} 
    && \big( 
    u^{a} \partial_{a} + u^{k}\partial_{k} \big)\,\rho^i= u^{i} =f^i
    \,\rho^i\, \quad 
    \rightarrow 
     \quad \{u^a,u^i\}=\{ 
        u^a,\rho^i f^i
        \}\,,
\end{eqnarray}
which follows from $\partial_k\rho^i=\delta_k^i$. The $i$ labels are not summed
over in the above equation.  I use the short-hand notation $\partial_a:=\partial_{\alpha^a}$ and
$\partial_k:=\partial_{\rho^k}$.  Additional constraints arise from requiring
that $\{u^a,f^i\rho^i\}$ give polynomial vector fields in loop-momentum variables. I
will deal with this constraint below in \sect{LoopMomParam}.

The physical meaning of the adapted coordinates suggests to introduce the
following notation for \ibp-generating vectors:
\begin{enumerate}
\item Horizontal vectors $\{u^a,u^i\}=\{ u^a,0\}$ inducing translations along
    slices of momentum space with fixed-propagator values.
\item Scaling vectors $\{u^a,u^i\}=\{ f_s\alpha^a ,\rho^i f^i \}$ inducing
    scaling transformations on the on-shell phase spaces. The corresponding
    \ibp-relations depend on the dimension parameter D.
\item Vertical vectors $\{u^a,u^i\}=\{ 0 ,\rho^i f^i \}$ which vanish on the
    maximal cut, and point vertical to slices of fixed propagator values.
\end{enumerate}
For the discussion of integrals with generic mass assignments the horizontal
vectors are sufficient for constructing surface terms. I focus on this
type below.

\section{Coordinates adapted to integral topology}
\label{LoopMomParam}

The general coordinate changes to coordinates adapted to integral
topologies is obtained in two steps \cite{IntDec}.
First for each strand in a multi-loop diagram a basis~\cite{NVbasis} of transverse
vectors $n^a$, $(n^a,p_j)=0$ and dual vectors $v^i=(G^{-1})^{ij} p_j$ is
introduced using the Gram matrix $G_{ij}=(p_i,p_j)$.
A distinct such basis is used per strand; two or three for two-loop diagrams. In
a second step the basis change to inverse propagators is setup,
\bea \label{LoopMom} \ell(\rho,\alpha):&=&\sum_i r_i\, v^i+\sum_a\alpha^a n^a
\,, \quad 
r_i:= -\frac{1}{2}\left(
(\rho^i+m_i^2-q_i^2)-(\rho^{i-1}+m_{i-1}^2-q_{i-1}^2)\right)\,,\\ && q_i=
\sum_{j=1}^i p_j\,,\quad c(\rho,\alpha):=  (\ell^2-m_0^2) - \rho^0=0.  \eea
It is helpful not to eliminate an $\alpha$ variable using the constraint
$c(\rho,\alpha)=0$. Instead one uses the redundant set of variables
$\{\alpha^a,\rho^i\}$ and keeps the quadratic constraint $c(\alpha,\rho)\sim0$
explicit. For multiple strands momentum conservation is imposed when strands
join in a vertex. 
In summary, at one-loop level one has one constraint, while one introduces three constraints
for planar two-loop integrals after solving for momentum
conservation~\cite{IntDec}.  Below multiple constraints are labeled by an index;
$c_i(\alpha,\rho)$. 

Numerator insertions are polynomial functions in the loop
momenta. It is an important property of the introduced coordinates, that
numerators are again polynomials in $\alpha^a$ and $\rho^i$ variables,
\begin{eqnarray}\label{tensorbasis}
(t_{\mu_1...  \nu_k}\ell^{\mu_1} ... \,\tell^{\nu_k}) \quad
\longleftrightarrow\quad 
\prod_{a,\ta,l,\tl} (\alpha^a)^{k_a} \,(\talpha^{\ta})^{k_{\ta}}\times
(\rho^l)^{k_l}(\trho^{\tl})^{k_{\tl}}\,.
\end{eqnarray}

The variables $k_i$ denote non-negative integers.  Polynomials that differ by a
multiple of the constraints $c_i(\alpha,\rho)$ give identical loop-momentum
functions as can be verified using eqn.~(\ref{LoopMom}). 

Polynomial surface terms are obtained from polynomial vector fields in momentum
space. Their defining property is that their directional derivatives map
polynomials to polynomials. One obtains two types of
restrictions; (1) when acting on the coordinate functions $\alpha^a$ and $\rho^i$
one finds that the components $\{f^i\rho^i,u^a\}$ are polynomials in $\alpha^a$
and $\rho^i$. (2) The equivalence of polynomials modulo the constraints is
maintained only if the vector fields map the constraint ideal to itself, 
\bea \label{defpolyvector}
\big(\sum_a u^a\partial_a+\sum_if^i\rho^i \partial_i\big) c_j(\alpha,\rho)\sim c_k(\alpha,\rho)\,.
\eea

Geometrically one reaches the same conclusion; the constraints
$c_i(\alpha,\rho)\sim 0$ state, that the physical momentum space is a sub
manifold ($c_i(\alpha,\rho)=0$) in the $\{\alpha,\rho\}$-coordinate space.
Vectors in loop-momentum space are naturally tangent vectors of this sub
manifolds and thus give zero when acting on the constraints $c_i(\alpha,\rho)$
on the solution space of $c_i(\alpha,\rho)=0$ consistent with
eqn.~(\ref{defpolyvector}).

\section{Explicit surface terms}
\label{offshellconstruction}
The surface terms are obtained from total derivatives of \ibp-generating
vectors.  The master integrals are obtained as numerators
in the complement of the surface terms.
This construction differs from the one at one-loop
level~\cite{OPP,NumUnitarity,RevUnitarity} which relied on tensor algebra and
symmetries of one-loop integrals, but reproduces the results. 
All \ibp-generating vectors are obtained from a
set of primitive ones, which I list in the following. By inspection one can
verify that the below vectors are horizontal \ibp-generating vectors.

The one-loop \ibp-generating vectors are given in standard notation by,
\begin{eqnarray}
    u_{[kl]}= (\ell,n_{[k|})(n_{|l]},\partial) \,.
\end{eqnarray}

The two-loop vectors are given by the following three types:
\begin{eqnarray}
\mbox{(a)} &&
    u_{[ijk]}=  (\hell,n_{[i|})(\ell,n_{|j|})(n_{|k]},\partial) \,,\quad\quad
\mbox{(b)} \quad 
    u_{[kl]}= (\ell,n_{[k|})(n_{|l]},\partial)+
    (\tell,\tn_{[k|})(\tn_{|l]},\tilde\partial)   \,,\\
   \mbox{(c)} && u_{[ij][kl]}= (\tell,\tn_{[k|})(\tn_{|l]},\hell) \,
    (\ell,n_{[i|})(n_{|j]},\partial)-
    (\ell,n_{[i|})(n_{|j]},\hell) \,
    (\tell,\tn_{[k|})(\tn_{|l]},\tilde\partial) \,.
\end{eqnarray}
The vectors generate (a) rotations around an axis, (b) diagonal rotations and
(c) crossed rotations, respectively. Tilde's indicate variables of a second
sub-loop.

Multiplying the above vectors with generic
tensors gives further \ibp{} vectors.
The surface terms are obtained by computing the divergence (\ref{ibpsurfaceterms}) with
\ibp{} vectors inserted. The numerators of the horizontal \ibp vectors are given by
\begin{eqnarray} \label{mainres}
\prod_{i,j} (n_i,\ell)^{k_i} \,(\tn_{j},\tell)^{k_{j}} \times u \quad \rightarrow\quad 
     \hm_{(u,{\vec k})} = 
    \big( u^\mu\partial_\mu + \tu^\nu\tpartial_\nu\big) \, \left(
    \prod_{i,j} (n_a,\ell)^{k_i} \,(\tn_{\ta},\tell)^{k_{j}} \right)\,.
\end{eqnarray}
since the divergence of the horizontal vectors vanishes and
they annihilate the propagators. The numerators $\hm_u$ (\ref{mainres}) give
the surface terms (\ref{surfaceterms}). This is true even if the powers
of the propagators altered.

\section{Total derivatives and master-integrand count}
\label{totdiffpullback}

One requires a complete set of tensor insertions for a given integral
topology. Systematic constructions of such
a basis of tensor insertions can be found at one-loop level in in
ref.~\cite{NumUnitarity} (see also \cite{RevUnitarity}) and for multi-loop
topologies in refs.~\cite{MLoopParamBFZ, MLoopAlgGeo, MLoopParamMO}.

For a given integral topology inserting an inverse propagator allows one to
cancel a propagator and leads to a reduced topology. Thus one considers
numerator tensors modulo inverse propagators ($\rho^i\sim0$).  Comparing to
(\ref{tensorbasis}) one identifies the independent numerator as the
polynomials,
\begin{eqnarray}\label{eqn:numbasis}
\prod_{a,\ta} (\alpha^a)^{k_a} \,(\talpha^{\ta})^{k_{\ta}} = \prod_{a,\ta}
(n_a,\ell)^{k_a} \,(\tn_{\ta},\tell)^{k_{\ta}}\,,
\end{eqnarray}
modulo the relations $c_i\sim0$ and $\rho^j\sim 0$.

These equivalence relations amount to imposing the
on-shell conditions, with all inverse propagators set to zero.  Thus, linear
independent numerator insertions are identified with independent functions on
the maximal-cut phase spaces. In order to count master integrands one works on
the unitarity cuts.

Cutting propagators amounts to
replacing the propagators with delta-distributions,
\begin{eqnarray}
    \int \frac{\rhomeasure}{\invprops} \,\times\,
\tm\,\mu\cutmeasure   \quad \cut\quad \int \tm(0,\alpha)
\,\mu|_{\rho^i=0} \cutmeasure\,.
\end{eqnarray}
Terms with some of the cut propagators missing are omitted in the cutting
prescription. 

For the \ibp-generating vectors ($u^i=f^i\rho^i$) cutting total derivatives
gives total derivatives on the unitarity cut surface,
\begin{eqnarray}\label{onoffmap}
\int \left[ \frac{ \partial_i \big( f^i \, \mu\, \tm \big)
}{\dropinvprops }  + \partial_b \left( \frac{u^b\, \mu\,  \tm }{\invprops }
\right)\right]  \measure
    \cut\quad   \int \partial_b \Big(u^b\, \mu\,  \tm   \Big) \cutmeasure\,.
\end{eqnarray}
The special property that the \ibp-generating vector fields are tangent vectors
of the maximal-cut phase spaces leads to this equation.

The number of master integrands is
given by the number of independent tensor insertions modulo the number of
surface terms. On-shell this amounts to the number of closed holomorphic forms
modulo exact holomorphic forms (total
derivatives), which is a topological index of the phase spaces.
Thus a topological property, in fact the number of half-maximal cycles of the
unitarity cut variety, counts master integrands.

\section{Summary and future directions}

I discussed new methods for the computation of multi-loop amplitudes
using a numerical unitarity approach.
The methods form a synthesis of established techniques to obtain integral
relations (or surface terms) \cite{KosowerIBP} and the unitarity approach. 

I presented a number of results from ref.\cite{IntDec}: First of all, I setup
the numerical unitarity approach for multi-loop amplitudes. I suggested
a classification of integral relations and constructed a complete set of
surface terms (\ref{mainres}) for massive planar two-loop topologies.
The surface terms are obtained from a new type of horizontal \ibp-generating
vectors. 
In addition, I pointed out the relation between the number of master integrands
and topological properties of the unitarity-cut phase spaces.
A central result is the geometric interpretation of \ibp-generating vectors and
convenient coordinates for dealing with multi-loop integrals. These methods
appear to be a useful extension to the mathematical toolbox for loop computations.

The best combination of analytic and numerical approaches for multi-loop
computations is not obvious in the moment, however, motivated by the one-loop
successes of the unitarity method it seems worth while to follow a similar
strategy for multi-loop computations.  I am optimistic that the new methods
discussed can contribute to precision predictions for the LHC experiments on
the long run.

\vspace{-0.4cm}
\acknowledgments
\vspace{-0.4cm}
This work is supported by a Marie Sk{\l}odowska-Curie Action Career-Integration
Grant PCIG12-GA-2012-334228 of the European Union and by the Juniorprofessor
Program of Ministry of Science, Research and the Arts of the state of
Baden-W\"urttemberg, Germany.

\end{document}